\documentclass [aps,showpacs,showkeys,preprint]{revtex4}
\usepackage{graphicx}
\usepackage{amsmath}

\begin{document}

\title{ \bf  Effects of interfacial pseudo-spin coupling
fluctuations on the dielectric properties of ferroelectric
superlattices }

\author{\bf Yin-Zhong Wu$^{1,2}$, Wei-Min Zhang $^{1}$, Ming-Rong Shen$^{3}$, and Zhen-Ya
Li$^{3}$}

\affiliation{$^{1}$Department of Physics, National Cheng Kung
University, Tainan 701,Taiwan\footnote{Email:yzwu@phys.ncku.edu.tw} \\
$^{2}$Department of Physics, Changshu Institute of Technology, Changshu 215500, China \\
$^{3}$Department of Physics, Suzhou(Soochow) University, Suzhou
215006, China}

\begin{abstract}
Using effective-field theory with correlations, we investigate the
effects of interfacial pseudo-spin coupling fluctuations on the
susceptibility and polarization of ferroelectric superlattices
within the framework of transverse Ising model. It is found that
the interfacial coupling fluctuations increase the susceptibility
in the low temperature region. For a strong interfacial coupling,
the phase transition temperature decreases with the strength of
fluctuations of the interfacial coupling. The dependence of the
susceptibility on the superlattice period of $BaTiO_{3}/SrTiO_{3}$
are plotted for different interfacial coupling fluctuations
strength. At room temperature, when the interfacial coupling
fluctuation increases, the peak position of the susceptibility
will shift to a large superlattice period.
\end{abstract}

\pacs{77.22.Ch; 77.55.+f}

\keywords{ Interfacial coupling fluctuation, ferroelectric
superlattice, dielectric property.}

\maketitle

\section{Introduction}

Recently developments in the fabrication of thin films have been
applied to the ferroelectric thin films. Due to their good
ferroelectric, dielectric, piezoelectric and pyroelectric
properties, ferroelectric heterostructures have attracted much
attention and have many potential applications, for instance, to
nonvolatile dynamic random access memories(FDRAM), thin film
capacitors, detectors, sensors, optical instruments. New
ferroelectric materials with excellent dielectric properties at
small sizes have also been considered to make electric device such
as the small size capacitors. Thus, a high dielectric constant
film with thickness less than $0.1 \mu m$ is a target for
ferroelectric research.

Giant permittivity associated with the motion of domain walls was
reported in epitaxial heterostructures having alternating layers
of ferroelectric and nonferroelectric oxides \cite{EKG}.
Experiments have also been carried out on the dielectric
enhancement and Maxwell-Wagner effects in ferroelectric
superlattice structure \cite{NR}. The lattice mismatch and
interfacial strain in ferroelectric multilayer are thought to be
the main causes for the dielectric enhancement \cite{NR,OBG}.
Theoretically, it is found that the dielectric property of the
ferroelectric superlattice is very sensitive to the interfacial
coupling and the thickness of the component \cite{QZP}. The
effects of stress on ferroelectric thin films have been studied
within the framework of Landau theory \cite{Zhu,WWZT} with the
conclusion that higher tensile stress enhances, while the higher
compressive stress reduces the mean susceptibility. By taking the
four-spin interaction into account, a ferroelectric-ferroelectric
phase transition was found for large periods at low temperature
\cite{WWZ}. In Refs.~\cite{YWL,WYDL}, the effects of long-range
interactions and a non-ferroelectric layer on the dielectric
properties of ferroelectric multilayer were studied. Defects zones
and the mismatch at the interface between successive ferroelectric
layers were observed in Ref.~\cite{OBG}. It is anticipated that a
ferroelectric multilayer with a high concentration of
interfaces(where the bonding and the structure will in general
depart from that of the interior of the layers or the bulk)will
have some new properties. A complex and inhomogeneous interface
may be important, or even dominant, in the superlattice structure
when the individual constituent layer is only a few unit cells. So
far, the effect of the interfacial structure on the dielectric
property of a ferroelectric superlattice has not been thoroughly
investigated. Analogous to the structure fluctuation or the bond
randomness in the amorphous ferromagnets \cite{BBKLM}, the
disordered structure of the ferroelectric interface must give rise
to the randomness in the interaction of the dipolar moments.

In this paper, we introduce interfacial pseudo-spin coupling
fluctuations in the transverse Ising model(TIM) to investigate the
effects of interfacial structure inhomogeneity on the dielectric
properties of ferroelectric superlattices. The coupling
fluctuations may also exist in the interior layers of the
superlattice. However, compared with the magnitude of the coupling
fluctuation within the interface, the fluctuations in the interior
layers are weaker. We focus our attention on the effects of the
interfacial pseudo-spin coupling fluctuation. The dielectric
property and phase transition temperature of ferroelectric Ising
superlattices have been investigated in the context of the
mean-field theory \cite{WS} and effective-field theory
\cite{ALKS}, and it is found that the dielectric constant has a
maximum for a small superlattice period at room temperature. It is
well known that mean-field theory should not be applied to
investigate fluctuation effects near the phase transition point.
Here we adopt effective-field theory (based on the Ising spin
identities and the differential operator technique) which is
superior to mean-field theory. It is found that in the low
temperature region the larger interfacial coupling fluctuations,
the higher the susceptibility of the superlattice. Remarkably,
interfacial coupling fluctuations lower the phase transition
temperatures when the interfacial coupling is stronger than the
pseudo-spin coupling in the interior layers. Finally, parameters
that imitate $BaTiO_{3}/SrTiO_{3}$ superlattice are applied to our
model. We find that large interfacial coupling fluctuations will
increase the dielectric constant of $BaTiO_{3}/SrTiO_{3}$
structure at room temperature, decrease the dielectric constant at
high temperature, and further more change the critical thickness
of the superlattice at which there exists a maximum value of the
dielectric constant. We postulate that an interfacial disorder
such as the interfacial coupling fluctuation is one of the reasons
for dielectric enhancement in the ferroelectric multilayer.

\newpage
\section{Model and Formulation}

We consider a ferroelectric superlattice composed of two different
components(A and B) with an interface (I). $L_{a}$, $L_{b}$, and
$L_{in}$ are the thicknesses respectively of A , B, and the
interface in a unit cell of the superlattice. Each layer is
defined on the x-y plane and with pseudo-spin sites on a square
lattice(see Fig.~1). As in Refs.~\cite{WS,ALKS}, we consider only
one interfacial layer($L_{in}=1$). The coupling constants between
pseudo spins in the interfacial layer and in the interior layer
can be different from that between spins within the interior
layers. The system is described by the Ising Hamiltonian with a
transverse field,
\begin{equation}
H=-\sum_{<ij>}J_{ij}S_{i}^{z}S_{j}^{z}-\sum_{<mn>}{\bar J}_{in}S_{m}^{z}S_{n}^{z}-\sum_{i}\Omega_{i}S_{i}^{x}-2\mu E\sum_{i}S_{i}^{z},
\end{equation}
where $\Omega_{i}$ is the transverse field. $S_{i}^{z}$,
$S_{i}^{x}$ are components of spin-1/2 operator at site $i$,
$J_{ij}$ is the coupling constant between the nearest neighbor
pseudo spins within the component A or B, ${\bar{J}}_{in}$ is the
nearest neighbor coupling constant in the interfacial layer and
that between the interface and the interior layers of component A
or B. $\bar{J}_{in}$ is assumed to be randomly distributed
according to the independent probability distribution function
$\rho ({\bar{J}}_{in})$.  $\mu$ is the effective dipole moment,
and $E$ is the applied electric field. The parameters $J_{ij}$ and
$\Omega_{i}$ are taken as:
\begin{equation}
J_{ij}=\left \{
\begin{array}{ll}
J_{a}&\ for \ i, j\in \ component\ A \\ J_{b}&\ for \ i, j\in \
component \ B,
\end{array}
\right.
\end{equation}

\begin{equation}
\Omega_{i}=\left \{
\begin{array}{ll}
\Omega_{a}&\ for \ i \in \ component\ A\\
\Omega_{in}&\ for \ i \in \ the \ interface \ I
\\\Omega_{b}&\ for \ i \in \ component \ B.
\end{array}
\right.
\end{equation}

For the ferroelectric material with the first-order phase
transition, a four-spin interaction term \cite{WWZ} must be
included in the Hamiltonian (1). Here, we will focus our attention
on the effects of the coupling fluctuation in the interface, and
the four-spin interaction term is not considered for simplicity.
The average values of pseudo spins in each layer of the
superlattice can be derived from the effective-field theory with
correlations. For instance, for $L_{a}=L_{b}=3$, we have
\begin{equation}
\begin{array}{lll}
R_{1}&=&<<S_{1}^{z}>>_{r}=<[cosh(\frac{1}{2}\bigtriangledown\bar{J}_{in})+2<S_{8}>sinh
(\frac{1}{2}\bigtriangledown\bar{J}_{in})].\\
&&[cosh(\frac{1}{2}\bigtriangledown
J_{a})+2<S_{1}>sinh(\frac{1}{2}\bigtriangledown J_{a})]^{4}.\\
&&[cosh(\frac{1}{2}\bigtriangledown
J_{a})+2<S_{2}>sinh(\frac{1}{2}\bigtriangledown
J_{a})]>_{r}.f(x,\Omega_{a})|_{x=0},\\
R_{2}&=&<S_{2}^{z}>=[cosh(\frac{1}{2}\bigtriangledown
J_{a})+2<S_{1}>sinh(\frac{1}{2}\bigtriangledown J_{a})].\\
&&[cosh(\frac{1}{2}\bigtriangledown J_{a})+2<S_{2}>sinh
(\frac{1}{2}\bigtriangledown J_{a})]^{4}.\\
&&[cosh(\frac{1}{2}\bigtriangledown J_{a})+2<S_{3}>sinh
(\frac{1}{2}\bigtriangledown
J_{a})].f(x,\Omega_{a})|_{x=0},\\
R_{3}&=&<<S_{3}^{z}>>_{r}=<[cosh(\frac{1}{2}\bigtriangledown
J_{a})+2<S_{2}>sinh(\frac{1}{2}\bigtriangledown J_{a})].\\
&&[\cosh(\frac{1}{2}\bigtriangledown
J_{a})+2<S_{3}>sinh(\frac{1}{2}\bigtriangledown
J_{a})]^{4}.\\
&&[cosh(\frac{1}{2}\bigtriangledown\bar{J}_{in})+2<S_{4}>sinh(\frac{1}{2}\bigtriangledown
\bar {J}_{in})]>_{r}.f(x,\Omega_{a})|_{x=0},\\
R_{4}&=&<<S_{4}^{z}>>_{r}=<[cosh(\frac{1}{2}\bigtriangledown\bar{J}_{in})+2<S_{3}>sinh(\frac{1}{2}\bigtriangledown\bar{J}_{in})].\\
&&[cosh(\frac{1}{2}\bigtriangledown\bar{J}_{in})+2<S_{4}>sinh(\frac{1}{2}\bigtriangledown\bar{J}_{in})]^{4}.\\
&&[cosh(\frac{1}{2}\bigtriangledown\bar{J}_{in})+2<S_{5}>sinh(\frac{1}{2}\bigtriangledown
\bar{J}_{in})]>_{r}.f(x,\Omega_{ab})|_{x=0},\\
R_{5}&=&<<S_{5}^{z}>>_{r}=<[cosh(\frac{1}{2}\bigtriangledown\bar{J}_{in})+2<S_{4}>sinh(\frac{1}{2}\bigtriangledown\bar{J}_{in})].\\
&&[cosh(\frac{1}{2}\bigtriangledown J_{b})+2<S_{5}>sinh
(\frac{1}{2}\bigtriangledown J_{b})]^{4}.\\
&&[cosh(\frac{1}{2}\bigtriangledown
J_{b})+2<S_{6}>sinh(\frac{1}{2}\bigtriangledown
J_{b})]>_{r}.f(x,\Omega_{b})|_{x=0},\\
R_{6}&=&<S_{6}^{z}>=[cosh(\frac{1}{2}\bigtriangledown
J_{b})+2<S_{5}>sinh(\frac{1}{2}\bigtriangledown J_{b})].\\
&&[cosh(\frac{1}{2}\bigtriangledown
J_{b})+2<S_{6}>sinh(\frac{1}{2}\bigtriangledown J_{b})]^{4}.\\
&&[cosh(\frac{1}{2}\bigtriangledown
J_{b})+2<S_{7}>sinh(\frac{1}{2}\bigtriangledown
J_{b})].f(x,\Omega_{b})|_{x=0},\\
R_{7}&=&<<S_{7}^{z}>>_{r}=<[cosh(\frac{1}{2}\bigtriangledown
J_{b})+2<S_{6}>sinh(\frac{1}{2}\bigtriangledown J_{b})].\\
&&[cosh(\frac{1}{2}\bigtriangledown
J_{b})+2<S_{7}>sinh(\frac{1}{2}\bigtriangledown J_{b})]^{4}.\\
&&[cosh(\frac{1}{2}\bigtriangledown\bar
{J}_{in})+2<S_{8}>sinh(\frac{1}{2}\bigtriangledown \bar
{J}_{in})]>_{r}.f(x,\Omega_{b})]|_{x=0},\\
R_{8}&=&<<S_{8}^{z}>>_{r}=<[cosh(\frac{1}{2}\bigtriangledown\bar{J}_{in})+2<S_{7}>sinh(\frac{1}{2}\bigtriangledown\bar{J}_{in})].\\
&&[cosh(\frac{1}{2}\bigtriangledown\bar{J}_{in})+2<S_{8}>sinh(\frac{1}{2}\bigtriangledown\bar{J}_{in})]^{4}.\\
&&[cosh(\frac{1}{2}\bigtriangledown\bar{J}_{in})+2<S_{1}>sinh(\frac{1}{2}\bigtriangledown\bar{J}_{in})]>_{r}.f(x,\Omega_{ab})|_{x=0},
\end{array}
\end{equation}

where function $f(x)$ is defined by
\begin{equation}
f(x,\Omega_{i})=\frac{x}{2\sqrt{x^{2}+\Omega
^{2}_{i}}}tanh(\frac{1}{2}\sqrt{x^{2}+\Omega ^{2}_{i}}).
\end{equation}

  To proceed further, we have to approximate the thermal multiple
correlations on the right side of Eq.~(4). We shall use the
Zernike decoupling approximation,
\begin{equation}
<S_{i}^{z}S_{j}^{z}...S_{k}^{z}S_{l}^{z}>\approx <S_{i}^{z}><S_{j}^{z}>...<S_{k}^{z}><S_{l}^{z}>.
\end{equation}

In order to describe the coupling fluctuations in a simple way, we
assume that the distribution of $J_{in}$ is taken to be
\begin{equation}
\rho ({\bar J}_{in})=\frac{1}{2}[\delta({\bar J}_{in}-J_{in}-\bigtriangleup J_{in})+\delta({\bar J}_{in}-J_{in}+\bigtriangleup J_{in})],
\end{equation}
and the parameter $\delta_{in}$ (which is introduced to describe
the magnitude of the coupling fluctuation in the interface) is
defined as
\begin{equation}
\delta_{in}=\frac{\bigtriangleup J_{in}}{J_{in}}.
\end{equation}

The symbol $<...>_{r}$ in Eq.~(4) denotes the average over random
bonds. These random bond averages are given by
\begin{equation}
\begin{array}{lll}
<cosh(\bigtriangledown {\bar J}_{in})>_{r}&=&cosh(\bigtriangledown J_{in}\delta_{in})cosh(\bigtriangledown J_{in}),\\
<sinh(\bigtriangledown {\bar J}_{in})>_{r}&=&cosh(\bigtriangledown J_{in}\delta_{in})sinh(\bigtriangledown J_{in}).
\end{array}
\end{equation}

The equations for $R_{i}$ in (4) ,where $i$ runs over all layers
in one period of the superlattice,  form a set of nonlinear
simultaneous equations from which each $R_{i}$ can be calculated
numerically. The average polarization of the superlattice can then
be obtained as
\begin{equation}
P_{av}=\sum_{i=1}^{L} {2\mu R_{i}}/L,
\end{equation}
with $L=L_{a}+L_{b}+L_{in}$. When the applied electric field $E$
is varied, the average susceptibility of the superlattice is
obtained numerically from

\begin{equation}
\chi=\frac{\partial P}{\partial E}|_{E=0}.
\end{equation}

By changing the value of $\delta_{in}$, we can investigate the
effects of the interfacial coupling fluctuations on the
susceptibility and the polarization of the ferroelectric
superlattice. The numerical results and discussions are given in
Sec.~III.

\newpage
\section{Results and discussions}

Effects of interfacial coupling and the thickness on the
ferroelectric multilayer have been studied in detail
\cite{QZP,MXS}. Here, we focus our attention on the effects of the
fluctuations of the interfacial pseudo-spin coupling. We first fix
the thickness of the superlattice to investigate the above
effects. In Fig.~2 and Fig.~3, we take $L_{a}=L_{b}=5$,
$J_{a}=2J_{b}$, and $\Omega_{a}=\Omega_{b}=0.5J_{b}$, where
$J_{b}$ is taken as the unit of energy. Fig.~2 and Fig.~3 are
plotted for weak and strong interfacial pseudo-spin couplings
respectively. As shown in Fig.~2 ($J_{in}=\sqrt{J_{a}J_{b}}$), for
a weak interfacial pseudo-spin coupling, the fluctuations of the
interfacial coupling result in an increment of the susceptibility
of the ferroelectric superlattice only in the low temperature
region. The phase transition temperature is almost constant as the
interfacial coupling fluctuation is increased. This is reasonable,
because the phase transition temperature is mainly determined by
the component $A$ which has a stronger pseudo-spin coupling than
in $B$ and in the interface. The average polarization of the
superlattice has a slight decrease as the interfacial coupling
fluctuation is increased (See Fig.~2(b)). The effects of the
interfacial coupling fluctuations are more pronounced for a strong
interfacial coupling ($J_{in}=3\sqrt{J_{a}J_{b}}$). In Fig.~3(a),
the susceptibility of the superlattice will increase greatly with
increasing $\delta_{in}$ below the transition temperature, and the
peak positions of the susceptibility shift to lower temperatures.
The dependence of the superlattce polarization on the temperature
for selected values of $\delta_{in}$ are plotted in Fig.~3(b). We
can thus see that the interfacial coupling fluctuations play an
important role in the dielectric properties and phase transition
temperatures of ferroelectric superlattices for strong interfacial
pseudo-spin couplings.

In Fig.~4, we fix the interfacial coupling fluctuation to
investigate the effects of the superlattice period on the
dielectric properties. With the increase of the period of the
superlattice, it is found that the peak value of the
susceptibility decreases and the peak position shifts to higher
temperatures. And a ferroelectric-ferroelectric phase transition
occurs for a large period of the superlattice at low temperature,
which is also observed by use of the mean-field theory in
Ref.~\cite{WWZ}.

In order to study the interfacial coupling fluctuation effects of
real physical systems, we consider a $BaTiO_{3}/SrTiO_{s}$
superlattice with the parameters in Fig.~1 chosen as
\cite{WS,ALKS} $J_{a}=264K$, $\Omega_{a}=0.01K$, $J_{b}=24K$,
$\Omega_{b}=87K$, $J_{in}=\sqrt{J_{a}J_{b}}$, and
$\Omega_{in}=\sqrt{\Omega_{a}\Omega_{b}}$. We also assume that the
interfacial coupling fluctuations are described by Eq.~(7). The
susceptibility curves are plotted in Fig.~5. When $\delta_{in}=0$,
there exists a peak value of the susceptibility around
$L_{a}=L_{b}=4$ at room temperature, and our result at
$\delta_{in}=0$ recovers that of Ref.~\cite{ALKS}. For large
fluctuations of the interfacial coupling ($\delta_{in}=6.0$), the
peak position of the susceptibility occurs at a large value of the
period at room temperature. However, as shown in Fig.~5(b), the
susceptibility at higher temperatures will decease with the
increase of $\delta_{in}$.

In summary, the susceptibility (and its peak value) of the
superlattice will increase with greater interfacial coupling
fluctuations in the low temperature region. When the interfacial
coupling is strong, the effects of the interfacial fluctuations on
the susceptibility and polarization are more pronounced than for
weak interfacial couplings. The peak position of the
susceptibility shifts towards a large period for large interfacial
coupling fluctuations at room temperature. We conclude that
interfacial coupling fluctuations is one of the reasons for
dielectric enhancement in the ferroelectric multilayer structure.

\begin{acknowledgments}
This work was supported by the National Science Council of ROC
under Contract Nos.~NSC-92-2112-M-006-024, NSC-92-2112-M-006-004
and the National Science Foundation of China under the grant
No.~10204016.
\end{acknowledgments}

\newpage

\begin{figure}[ht]
   \includegraphics*[width=4in]{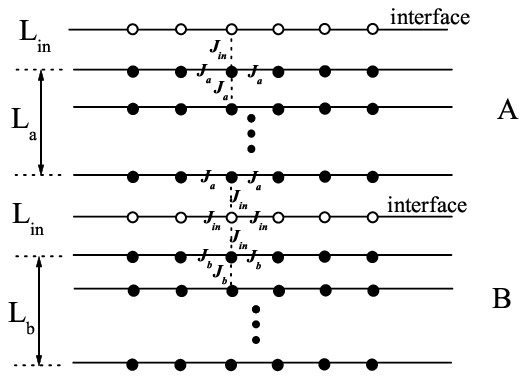}
    \caption{Schematic illustration of the model of a ferroelectric
superlattice. The interfacial layer is marked by hollow symbols.}
\end{figure}

\newpage
\begin{figure}[ht]
   \centering
   \includegraphics[width=4in]{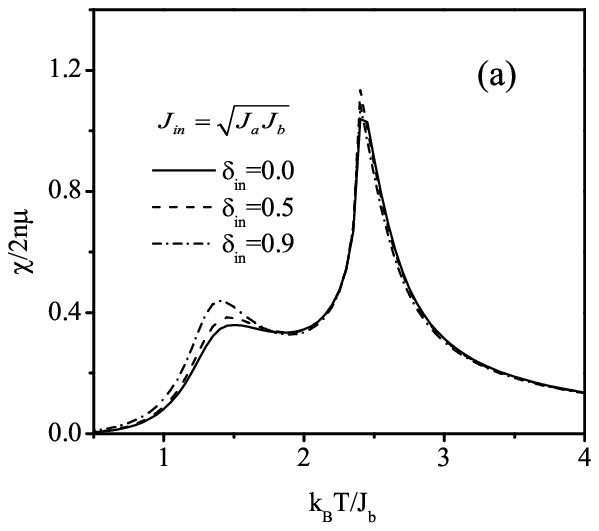}
   \includegraphics[width=4in]{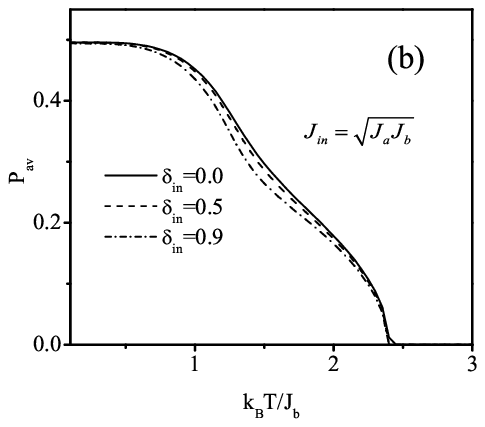}
   \caption{The dependence of the susceptibility(a) and the mean
polarization(b) of the ferroelectric superlattice on the
temperature for a given weak interfacial coupling $J_{in}$ and
different interfacial coupling fluctuations $\delta_{in}$.}
\end{figure}

\begin{figure}[ht]
   \centering
   \includegraphics[width=4in]{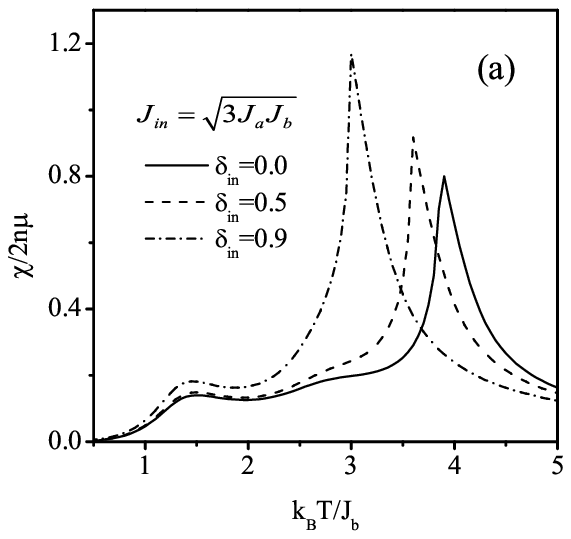}
   \includegraphics[width=4in]{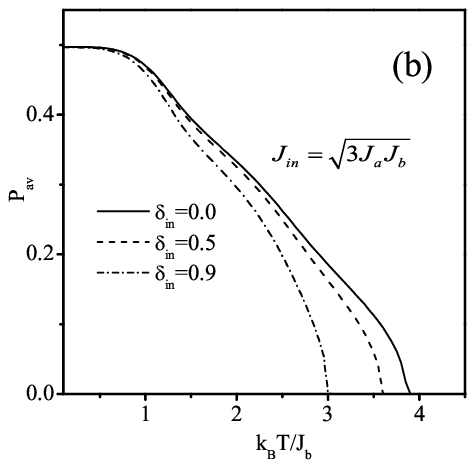}
   \caption{The dependence of the susceptibility(a) and the mean
polarization(b) of the ferroelectric superlattice on the
temperature for a given strong interfacial coupling and different
interfacial coupling fluctuations $\delta_{in}$.}
\end{figure}

\newpage
\begin{figure}[ht]
   \centering
   \includegraphics[width=4in]{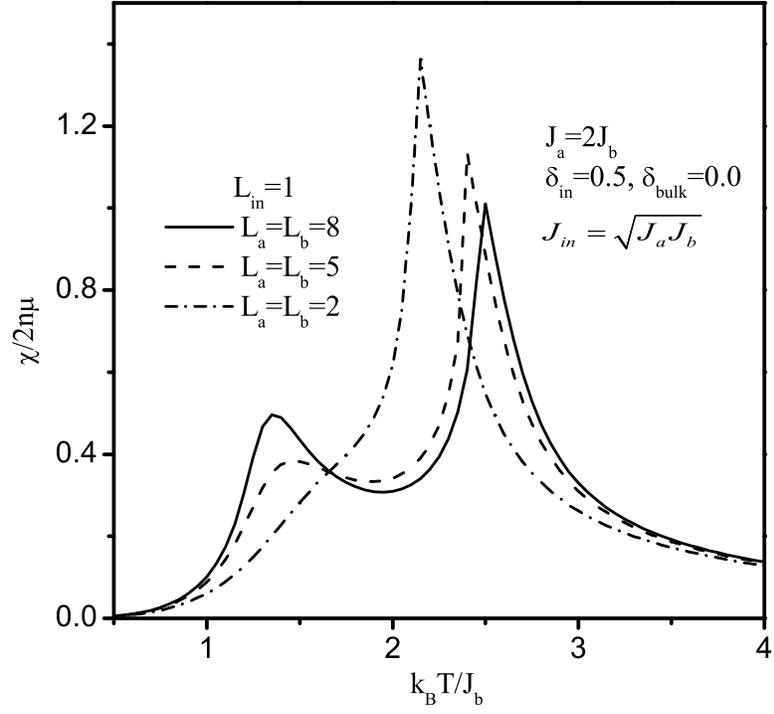}
   \caption{Plot of the susceptibility versus temperature of the ferroelectric
superlattice with different superlattice periods.}
\end{figure}

\begin{figure}[ht]
   \centering
   \includegraphics[width=4in]{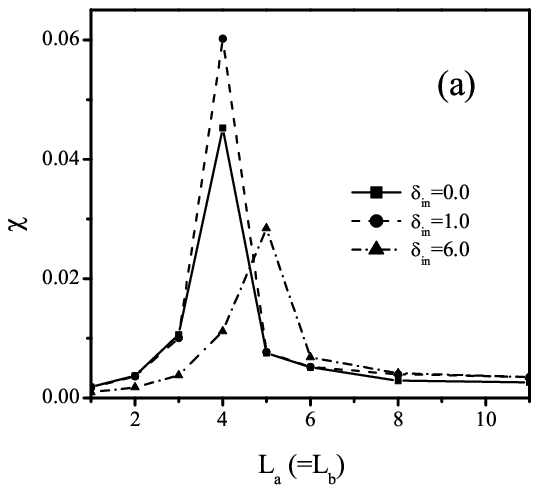}
   \includegraphics[width=4in]{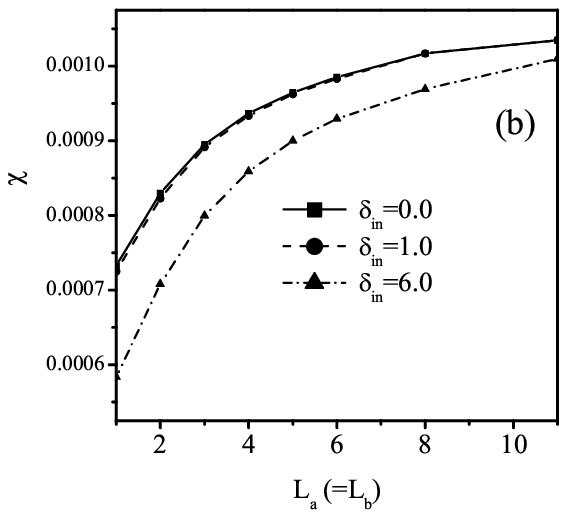}
   \caption{The period dependence of susceptibility for some selected
interfacial coupling fluctuations (a) at room temperature, (b) at
500 K.}
\end{figure}

\end{document}